\documentclass[aip,jcp,preprint,nofootinbib]{revtex4-1}

\usepackage[utf8]{inputenc}
\usepackage[margin=2.5cm]{geometry}
\usepackage{amsmath}
\usepackage{graphicx}
\usepackage{tabularx}
\usepackage[breaklinks=true, hypertexnames=false]{hyperref}
\usepackage{amsfonts}
\usepackage{amssymb}
\usepackage{mathtools}
\DeclarePairedDelimiter\ceil{\lceil}{\rceil}

\usepackage{natbib}
\usepackage{amsmath}
\usepackage{chemarrow}
\usepackage{color}
\usepackage{float}
\usepackage{enumerate}
\usepackage{bibentry}
\nobibliography*

\usepackage{soul}

\begin{document}

\title{Multi-level methods and approximating distribution functions}


\author{D. Wilson}
\email{daniel.wilson@dtc.ox.ac.uk}
\affiliation{$^{1)}$Mathematical Institute, University of Oxford, Radcliffe Observatory Quarter, Woodstock Road, Oxford, OX2 6GG}
\author{R.E. Baker}
\affiliation{$^{1)}$Mathematical Institute, University of Oxford, Radcliffe Observatory Quarter, Woodstock Road, Oxford, OX2 6GG}

\begin{abstract}
Biochemical reaction networks are often modelled using discrete-state, continuous-time Markov chains. System statistics of these Markov chains usually cannot be calculated analytically and therefore estimates must be generated via simulation techniques. There is a well documented class of simulation techniques known as exact stochastic simulation algorithms, an example of which is Gillespie's direct method. These algorithms often come with high computational costs, therefore approximate stochastic simulation algorithms such as the tau-leap method are used. However, in order to minimise the bias in the estimates generated using them, a relatively small value of tau is needed, rendering the computational costs comparable to Gillespie's direct method.

The multi-level Monte Carlo method (Anderson and Higham, Multiscale Model. Simul. 10:146--179, 2012) provides a reduction in computational costs whilst minimising or even eliminating the bias in the estimates of system statistics. This is achieved by first crudely approximating required statistics with many sample paths of low accuracy. Then  correction terms are added until a required level of accuracy is reached. Recent literature has primarily focussed on implementing the multi-level method efficiently to estimate a single system statistic. However, it is clearly also of interest to be able to approximate entire probability distributions of species counts. We present two novel methods that combine known techniques for distribution reconstruction with the multi-level method. We demonstrate the potential of our methods using a number of examples.
\end{abstract}

\maketitle


\section{Introduction}

Biochemical reaction networks describe interactions between the molecular populations of biological and physiological systems. A classical  deterministic approach to modelling these networks is to use systems of ordinary or partial differential equations to describe the evolution of the concentrations of each species. However, experimental researchers, such as Elowitz et al. \cite{fluorescence}, have shown that stochasticity can be observed in a variety of biological phenomena including gene expression within the cell. Thus, deterministic modelling can neglect important characteristics of a system. For example, in systems with low molecular populations using a deterministic model to describe the molecular concentrations can fail to predict a phenomenon known as stochastic focussing in signalling networks \cite{Szekely_2012}. Deterministic approaches can also fail to accurately replicate the dynamics of systems with large molecular numbers, such as systems operating near bifurcation points \cite{SNIPER}. 

To include stochasticity into models of biochemical reaction networks one can consider the framework of the chemical master equation \cite{Jahnke}. This assumes spatial homogeneity in the system, which is the case we shall consider in this paper. The chemical master equation describes the temporal evolution of molecular species in terms of the number of molecules present rather than their concentrations. This description provides a system of ordinary differential equations (ODEs, one for each possible state of the system) that, for simple cases involving only zeroth and first order reactions, can be solved analytically \cite{Jahnke}. For more complicated systems that include higher order reactions analytical solutions are not generally feasible so one may try and use a numerical scheme to approximate the solution to the chemical master equation \cite{Engblom,Jahnke_2,Jahnke}. However, this is also often infeasible due to the high (possibly infinite) number of distinct states in the system, so the only option to explore system behaviour is the use of simulation techniques.

There is a class of simulation techniques known as exact stochastic simulation algorithms (eSSAs) which model systems precisely, in the sense that the sample paths generated using them are mathematically equivalent to the chemical master equation. However, the more complex the system, the longer these simulations take to run, and for practical purposes approximate stochastic simulation algorithms (aSSAs) that favour speed over accuracy are often used. Unfortunately, obtaining accurate summary statistics for a system using aSSAs often renders the computational costs of those algorithms on a par with those of eSSAs. 

Recently proposed by Anderson and Higham \citep{Anderson_Higham_12} is the discrete-state multi-level method. This method vastly reduces the computational cost associated with generating summary statistics for a given system compared with traditional eSSAs and aSSAs. It does so by generating many sample paths of poor accuracy with very little computational effort and then reduces the error of summary statistics generated using these sample paths by adding correction terms from a smaller number of sample paths of higher accuracy. The paper by Lester et al. \cite{userguide} provides a clear introduction to the multi-level method with the focus being on implementation. In this work, we show how to exploit the multi-level method to accurately and efficiently estimate cumulative distribution functions (CDFs) of molecular species at a terminal time, $T$.

In Section \ref{section:modelling} we briefly describe the basic simulation techniques that will be used throughout the paper, namely Gillespie's direct method \cite{Gillespie} (Gillespie's DM) and the tau-leap method \cite{tau-leap}. We also introduce the multi-level method, stating how to optimally choose the number of sample paths on each level and simulate these paths with low sample variance. Section \ref{section:distribution} focusses on two methods of approximating the CDF of a chosen molecular species at time $T$, where we use the Schl\"{o}gl system \cite{schlogl_origin} and a dimerisation model \cite{Anderson_Higham_12} as example cases. Then we combine these two approximation methods with the multi-level method. We investigate how to optimise implementation of our proposed methods, resulting in two adaptive algorithms designed to use the multi-level method efficiently without any prior knowledge of the system. We conclude, in Section \ref{section:discussion}, with a short discussion.


\section{Modelling and simulating biochemical reaction networks} \label{section:modelling}

Let $S_1 , \ldots , S_N$ represent $N$ distinct populations or species that may be involved in $M$ reaction channels, $R_1 , \ldots , R_M$, in a system with volume $\nu$. We assume that the system is well stirred in order that we may ignore any spatial effects in our model. Denote the size of population $S_i$ at time $t$ as $X_i(t)$. We define the state space vector at time $t$ as
\begin{equation}
 \textbf{X}(t) = (X_1(t) , \ldots , X_N(t))^T.
\end{equation}

We also introduce the stoichiometric or state-change matrix, $\boldsymbol \nu = \{\nu_{ij}\}$, where $\nu_{ij}$ is the change in the copy number of $S_i$ after the reaction channel $R_j$ fires. Thus $\boldsymbol \nu$ is an $N \times M$ matrix. If reaction $R_j$ fires at time $t+s$ given no other reactions in $[t,t+s)$, we update the state space vector as follows
\begin{equation}
 \textbf{X}(t+s) = \textbf{X}(t) + \boldsymbol\nu_j,
\end{equation}
where $\boldsymbol \nu_j$ is the stoichiometric vector, defined as the $j^{th}$ column of the stoichiometric matrix,
\begin{equation}
\boldsymbol\nu_j = (\nu_{1j} , \ldots , \nu_{Nj})^T.
\end{equation}

Each reaction $R_j$ has a propensity function $a_j(\textbf{X}(t))$ at time $t$, this quantity determines the rate at which the $j^{th}$ reaction occurs. The propensity functions, $a_j(\textbf{X}(t))$, can be calculated by considering the number of ways in which a reaction can occur given the copy numbers at time $t$. A detailed description can be found in the paper by Erban et al.\cite{Radek-guide}.

The Kurtz representation \cite{Kurtz-2011} allows us to describe evolution of the state space vector as follows:
\begin{equation} \label{equation:statespace}
\textbf{X}(t) = \textbf{X}(0) + \sum_{j=1}^M Y_1^j \left( \int_0^t a_j(\textbf{X}(s)) \, \textrm{d}s \right) \cdot \boldsymbol \nu_j,
\end{equation}
where the $Y_1^j$, $j = 1 , \ldots , M$, are Poisson processes of unit rate.

\subsection{Gillespie's direct method}

Gillespie's DM \cite{Gillespie} is a standard example of an eSSA for generating sample paths from (\ref{equation:statespace}) We include a step by step algorithm below:

\begin{enumerate}
\item Initialise the copy numbers, $\textbf{X}(0)$, and the stoichiometric matrix, $\boldsymbol \nu$. Choose a terminal time, $T$, and set $t=0$.

\item Calculate the propensity functions, $a_j=a_j(\textbf{X}(t))$, for each reaction channel, $j$, and $a_0 = \sum_{j=1}^M a_j$. 

\item Generate $r_1 \sim \mathbf{U}\left(0,1\right)$ and set $\Delta = 1/a_0 \mathrm{log}(1/r_1)$.

\item If $t+\Delta > T$ then terminate the algorithm, otherwise continue to step $5$.

\item Choose the next reaction to occur by finding the minimal $k$ such that $\sum_{j=1}^k a_j > a_0 \times r_2$, where $r_2 \sim \mathbf{U}\left(0,1\right)$. 

\item Update the copy numbers, $\textbf{X}(t+\Delta) = \textbf{X}(t) + \boldsymbol \nu_k$.

\item Let $t = t + \Delta$ and return to step $2$. 
\end{enumerate}

The main advantage of this algorithm is that it can be used to produce unbiased estimates of system statistics at a terminal time, $T$. As an example, for the mean copy number of $S_i$ we generate $n$ sample paths to calculate
\begin{equation}
\mathbb{E}(X_i(T)) = \dfrac{1}{n}\sum_{r=1}^n X_i^{(r)}(T),
\end{equation}
where $X_i^{(r)}$ represents the copy number of $S_i$ from the $r^{th}$ sample path. The disadvantage of this eSSA is that it simulates every reaction individually, which can render it very computationally expensive. In order to reduce computation time, we now introduce a more efficient algorithm, known as the tau-leap method, which provides sample paths that approximately follow (\ref{equation:statespace}).

\subsection{The tau-leap method}

The tau-leap method is an aSSA, first proposed by Gillespie \cite{tau-leap}. It is designed to reduce computation time by temporarily fixing the propensity functions and firing several reactions during each time interval of length $\tau$. The method is equivalent to taking a Forward Euler approximation \cite{suli} to equation (\ref{equation:statespace}),
\begin{equation}
\textbf{X}(t+\tau) = \textbf{X}(t) + \sum_{j=1}^M Y_1^j\left(a_j\left(\textbf{X}\left(t\right)\right) \cdot \tau\right) \cdot \boldsymbol \nu_j.
\end{equation}
A step by step algorithm is again presented below:
\begin{enumerate}
\item Initialise the copy numbers, $\textbf{X}(0)$, and the stoichiometric matrix, $\boldsymbol \nu$. Choose a terminal time, $T$, and a time step, $\tau$, such that $T/\tau$ is an integer. Set $t=0$.

\item If $t+\tau > T$ then terminate the algorithm, otherwise continue to step $3$.

\item Calculate the propensity functions, $a_j = a_j(\textbf{X}(t))$, for each reaction channel, $j$.

\item Generate Poisson random variates, $p_j$, from $Y_1^j(a_j \cdot \tau)$ for $j=1, \ldots ,M$.

\item Update the copy numbers $\textbf{X}(t+\tau) = \textbf{X}(t) + \sum_{j=1}^M p_j \cdot \boldsymbol \nu_j$.

\item Let $t = t+\tau$ and return to step $2$.
\end{enumerate}

An attractive feature of the algorithm is that the accuracy of the approximation is determined by the choice of $\tau$. The smaller the time step, $\tau$, the greater the accuracy of the method, and indeed as $\tau \downarrow 0$ the method is equivalent to Gillespie's DM \cite{tau-leap}. However, often the value of $\tau$ needed to gain the required level of accuracy is small enough that the CPU time becomes comparable to that of Gillespie's DM, and so for further computational savings we utilise the multi-level Monte Carlo method \cite{Anderson_Higham_12, Giles}. 

\subsection{The multi-level Monte Carlo method}

In this section we provide a very brief introduction to the multi-level Monte Carlo method \cite{Anderson_Higham_12, Giles}. Heuristically, the method combines the best attributes of Gillespie's DM and the tau-leap algorithm, to produce fast and unbiased estimates of system statistics. Further information can be found in the work by Anderson and Higham \cite{Anderson_Higham_12} and Giles \cite{Giles}, as well as Lester et al. \cite{userguide}.

Suppose we are interested in calculating an estimate of $X_i(T)$. We start at the base level, otherwise known as level $0$. The estimate at this level is calculated using the tau-leap algorithm, as discussed in the previous section, with $\tau = \tau_0$. If $n_0$ samples are generated, then we have the biased estimate,
\begin{equation}
Q_0 = \mathbb{E}\left(Z_{\tau_0}\right) = \dfrac{1}{n_0} \sum_{r=1}^{n_0} Z_{\tau_0}^{(r)},
\end{equation}
where $Z_{\tau_0}^{(r)}$ is the copy number of population $i$ at time $T$ from the $r^{th}$ sample using the tau-leap algorithm with $\tau=\tau_0$. Typically, a large $\tau_0$ is chosen so that $Q_0$ can be calculated quickly. The purpose of the other levels is to reduce the bias of this crude approximation, $Q_0$, by adding correction terms.

The first level correction is calculated by generating two sets of $n_1$ sample paths using the tau-leap aSSA. One set has $\tau = \tau_0$ and the other $\tau = \tau_1 < \tau_0$. From the two sets of paths, we calculate the correction term as follows:
\begin{equation}
Q_1 = \mathbb{E}\left(Z_{\tau_1} - Z_{\tau_0}\right) \approx \dfrac{1}{n_1} \sum_{r=1}^{n_1} \left(Z_{\tau_1}^{(r)} - Z_{\tau_0}^{(r)}\right).
\end{equation}
This correction is then added to the initial estimate from the base level:
\begin{equation}
Q_0 + Q_1 = \mathbb{E}\left(Z_{\tau_0}\right) + \mathbb{E}\left(Z_{\tau_1} - Z_{\tau_0}\right) = \mathbb{E}\left(Z_{\tau_1}\right).
\end{equation}
This estimator has bias equivalent to just using the tau-leap aSSA with $\tau = \tau_1$. To further decrease the bias of the estimate we introduce time steps $\tau_{\ell} < \tau_{\ell-1} < \ldots < \tau_0$, and then add additional correction levels,
\begin{equation}
Q_\ell = \mathbb{E}\left(Z_{\tau_\ell} - Z_{\tau_{\ell-1}}\right) = \dfrac{1}{n_\ell} \sum_{r=1}^{n_\ell} \left( Z_{\tau_\ell}^{(r)} - Z_{\tau_{\ell-1}}^{(r)} \right),
\end{equation}
to give
\begin{equation}
 \mathbb{Q}_b \coloneqq \sum_{\ell=0}^L Q_\ell = \mathbb{E}\left(Z_{\tau_0}\right) + \sum_{\ell=1}^{L} \mathbb{E}\left(Z_{\tau_\ell} - Z_{\tau_{\ell-1}}\right) = \mathbb{E}\left(Z_{\tau_L}\right).
\end{equation}

This means that the estimator $\mathbb{Q}_b$ has bias equivalent to that generated using the tau-leap aSSA with $\tau = \tau_L$. The number of levels, $L$, can be chosen such that a desired level of accuracy is reached. However, a final correction level may be used in order to gain an unbiased estimate, if required. Again we generate two sets of paths, one set from a tau-leap algorithm where $\tau = \tau_L$ and one set from an eSSA such as Gillespie's DM, using $n_{L+1}$ samples. We can then calculate
\begin{equation}
Q_{L+1}^{\dag} = \mathbb{E}\left(X_i^{'} - Z_{\tau_L}\right) = \dfrac{1}{n_{L+1}} \sum_{r=1}^{n_{L+1}} \left(X_i^{'(r)}\left(T\right) - Z_{\tau_L}^{(r)}\right),
\end{equation}
where $X_{i}^{'}$ is an identical and independently distributed (i.i.d.) copy of $X_{i}$. This last correction can then be added to the previous estimate and, using the linearity of expectation, we have
\begin{equation}
\mathbb{E}\left(X_i\left(T\right)\right) = \mathbb{E}\left(Z_{\tau_0}\right) + \sum_{\ell=1}^{L} \mathbb{E}\left(Z_{\tau_\ell} - Z_{\tau_{\ell-1}}\right) + \mathbb{E}\left(X_i^{'} - Z_{\tau_L}\right) = \sum_{\ell=0}^L Q_\ell + Q_{L+1}^{\dag}.
\end{equation}

For the rest of this paper we shall choose $\tau_\ell = \tau_0/K^{\ell}$ where $K \in \mathbb{N}^{+}$. This means that the time steps used on consecutive levels are nested and renders the algorithm simple to implement. However, we note that it is not always the most appropriate choice: when dealing with systems that exhibit behaviours on a range of timescales a more sophisticated choice is necessary \cite{adaptive-lester}.

The reduction in computational cost for the multi-level method is achieved by introducing two coupling techniques. A tau-leap coupling technique for the sets of sample paths on the first $L$ levels, and an exact coupling technique for the $\left( L+1 \right)^{st}$ level. The coupling is designed to create a positive correlation between coarse, $Z_{\tau_{\ell-1}}^{(r)}$, and fine, $Z_{\tau_{\ell}}^{(r)}$, sample paths so that $\mathrm{Var}\left(Z_{\tau_{\ell}} - Z_{\tau_{\ell-1}} \right)$ is reduced and fewer sample paths need to be generated to obtain an estimator with a desired accuracy. The following algorithm is the tau-leap coupling technique:

\begin{enumerate}
\item Initialise the copy numbers for the coarse and fine paths, $\textbf{Z}_c(0)$ and $\textbf{Z}_f(0)$, and the stoichiometric matrix, $\boldsymbol \nu$. Choose a terminal time, $T$, base level leap size, $\tau_0$, and scaling factor, $K$. Set $t=0$.

\item For $\alpha = 1$ to $T/\tau_{\ell-1}$:
\begin{enumerate}

\item Calculate the propensity functions at the current time $t$ for the coarse system, $a_j^c = a_j(\textbf{Z}_c(t))$, for each reaction channel, $j$.

\item For $\beta = 1$ to $K$:

\begin{enumerate}

\item Calculate the propensity functions at the current time $t$ for the fine system, $a_j^f = a_j(\textbf{Z}_f(t))$, for each reaction channel, $j$.

\item Define the following three propensity functions:
\begin{gather}
b_j^1 = \mathrm{min}(a_j^c,a_j^f);\\
b_j^2 = a_j^c - b_j^1; \\
b_j^3 = a_j^f - b_j^1;
\end{gather}
for each reaction channel, $j$.

\item For $j=1,\ldots,M$ generate Poisson random variates, $Y_j^r$, each with rate $b_j^r \cdot \tau_\ell$ for $r \in \{1,2,3\}$.

\item Update the copy numbers of the two systems as follows:
\begin{equation}
\textbf{Z}_c(t+\tau_\ell) = \textbf{Z}_c(t) + \sum_{j=1}^M (Y_j^1 + Y_j^2) \cdot \boldsymbol \nu_j;
\end{equation}
\begin{equation}
\textbf{Z}_f(t+\tau_\ell) = \textbf{Z}_f(t) + \sum_{j=1}^M (Y_j^1 + Y_j^3) \cdot \boldsymbol \nu_j.
\end{equation}

Let $t = t + \tau_\ell$. 

\end{enumerate}
\end{enumerate}
\end{enumerate}
Further details, including those detailing the exact coupling technique, can be found in the paper by Lester et al. \cite{userguide}.

Finally, we need an algorithm to evaluate how many paths, $n_{\ell}$, to generate on each level, $\ell$, such that we minimise the expected CPU time of the method whilst controlling the variance, $\mathbb{V} = \sum_{\ell=0}^L V_{\ell}$ (the unbiased case can be considered analogously), where $V_{\ell}$ is the estimator variance on the $\ell^{th}$ level, of the summary statistic of interest (here the first moment $\mathbb{E}\left(X_i \right)$) to within a user prescribed tolerance, $\epsilon$.

If a pair of paths on the $\ell^{th}$ level takes $c_{\ell}$ units of CPU time to be generated, we wish to minimise
\begin{equation} \label{equation:CPU}
\sum_{\ell=0}^L n_{\ell} c_{\ell},
\end{equation}
subject to
\begin{equation}
\sum_{\ell=0}^L V_{\ell} < \epsilon.
\end{equation}
This optimisation problem can be solved using the method of Lagrange multipliers \cite{userguide} and suggests that we should take
\begin{equation} \label{equation:paths}
n_{\ell} = \ceil*{\dfrac{V_{\ell}/c_{\ell}}{\epsilon \sum_{\ell=0}^L \sqrt{c_{\ell}V_{\ell}}}},
\end{equation}
where $\ceil*{.}$ represents the ceiling function. One means to estimate $c_{\ell}$ and $V_{\ell}$ involves generating an initial number of sample paths on each level (usually $100$ or $1000$) \cite{Anderson_Higham_12}.

\section{Reconstructing cumulative distribution functions} \label{section:distribution}

In this section we introduce two methods to approximate CDFs of discrete state systems and then how to implement these methods efficiently alongside the multi-level method. We demonstrate the utility of our approach using two explanatory examples.

\subsection{Using a maximum entropy approach}

We first consider approximating the CDF of species $X_i$ at terminal time $T$ using a finite set of $M$ moments, $\{\mu_1,\ldots,\mu_M\}$, and the method of maximising entropy \cite{jaynes} to first estimate the corresponding probability distribution function (PDF). For a distribution on the discrete state space $\Omega = \{x_1,\ldots,x_m\}$ with PDF $\mathbf{p} = \left(p_1,\ldots,p_m\right)$, the entropy function, $\mathbb{H}(\mathbf{p})$, is defined as
\begin{equation} \label{equation:entropy}
\mathbb{H}(\mathbf{p}) \coloneqq \sum_{i=1}^m p_i \mathrm{log}\left(\dfrac{1}{p_i}\right).
\end{equation}
If $m=\infty$ we select an integer $N$ at which to truncate the state space, so that we work with $\Omega_t = \{x_1,\ldots,x_N\}$.
The optimisation problem is now to find the PDF, $\mathbf{p}$, which maximises the entropy function (\ref{equation:entropy}) subject to the moment constraints
\begin{equation}
\mu_j = \sum_{i=1}^N (x_i)^j p_i,
\end{equation}
for $j \in \{0,\ldots,M\}$. Note we include $\mu_0$, and set $\mu_0 = 1$ as we require $\sum_{i=1}^N p_i = 1$. The solution is obtained by introducing $M+1$ Lagrange multipliers \cite{Calculus}, $\boldsymbol \lambda = (\lambda_0,\ldots,\lambda_M)$, and the Lagrange functional
\begin{equation}
\Gamma = \sum_{i=1}^N p_i \mathrm{log}\bigg(\dfrac{1}{p_i}\bigg) - \sum_{j=1}^{M} \lambda_j\bigg(\sum_{i=1}^N (x_i)^j p_i - \mu_j\bigg) - c\bigg( \sum_{i=1}^N p_i - \mu_0 \bigg),
\end{equation}
where $c = \lambda_0 -1$. By solving $\partial \Gamma / \partial p_i = 0$, we obtain the solution
\begin{equation}
p_i = \mathrm{exp} \bigg( -\sum_{\ell=0}^M \lambda_{\ell} (x_i)^{\ell} \bigg),
\end{equation}
for $i \in \{1,\ldots,N\}$. Now that we know the form of the solution, which has $M+1$ unknowns, $\boldsymbol \lambda = (\lambda_0,\ldots,\lambda_M)$, we have a system of $M+1$ nonlinear equations,
\begin{equation}
\sum_{i=1}^N \mathrm{exp} \bigg( -\sum_{\ell=0}^M \lambda_{\ell} (x_i)^{\ell} \bigg)(x_i)^j = \mu_j,
\end{equation}
for $j \in \{0,\ldots,M\}$. We need to employ a numerical method to calculate $\boldsymbol \lambda = (\lambda_0,\ldots,\lambda_M)$, the one used in this paper is adapted from work by Mohammad-Djafari \cite{matlab-entropy}. The CDF can then be trivially calculated.

We now present possible ways of efficiently implementing this method in combination with the multi-level method in order to estimate CDFs of given species numbers. To aid us we use two example models, the Schl\"{o}gl model and a dimerisation model.

\subsubsection{The Schl\"{o}gl model}
\begin{figure}[H] 
\centering
$\begin{array}{cc}
\includegraphics[scale=1.1]{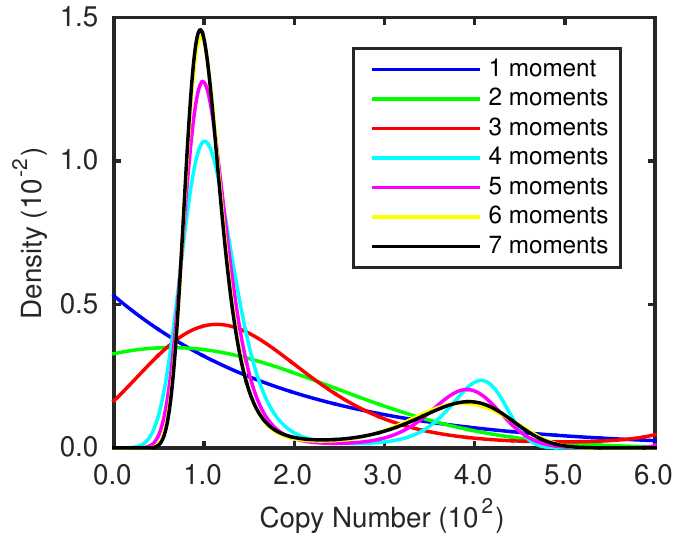} & \includegraphics[scale=1.1]{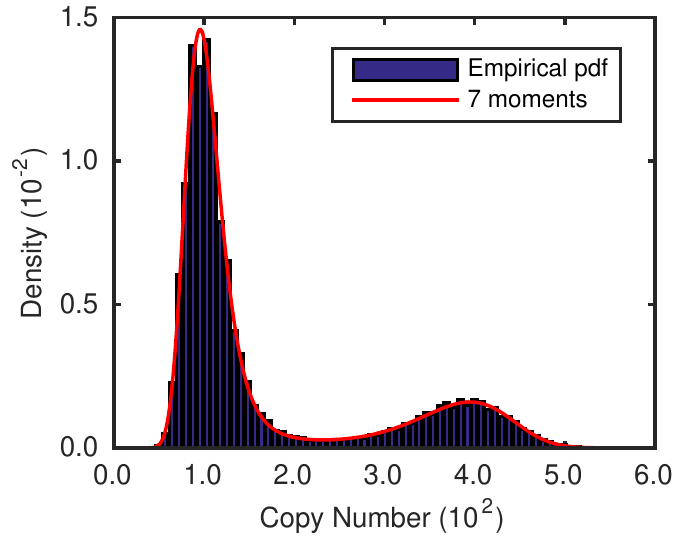}
\end{array}$
\caption[Maximum entropy PDFs using the first seven moments]{Left: PDFs from the maximum entropy method using up to the first seven moments on the truncated state space $\Omega_t = \{0,1,2,\ldots,600\}$. Right: empirical PDF and the PDF generated using the maximum entropy method and the first seven moments.}
\label{figure:optimalmoments}
\end{figure}

The Schl\"{o}gl model \cite{schlogl_origin,schlogl} consists of a single species, $A$, and four possible reactions:
\begin{equation} \label{equation:schlogl:reactions}
R_1 : \emptyset \xrightarrow{k_1} A;\hspace{10 mm} R_2 : A \xrightarrow{k_2} \emptyset;\hspace{10 mm} R_3 : 2A \xrightarrow{k_3} 3A;\hspace{10 mm} R_4 : 3A \xrightarrow{k_4} 2A. 
\end{equation}
We fix the rates to be $[k_1,k_2,k_3,k_4] = [2200, 37.5, 0.18, 0.00025]$ for the remainder of this paper. We chose the initial condition $A(0)=0$ and terminal time $T=20$. The first question we need to answer is how many moments are necessary to approximate the CDF of $A$ to within a desired degree of accuracy. To gain insight into this problem, we generated $10^6$ sample paths using Gillespie's DM and calculated the first seven moments. With these moments we calculated approximate PDFs for $A(20)$ using up to and including seven moments in the method of maximum entropy. The results are presented on the left of Figure \ref{figure:optimalmoments}.

Figure \ref{figure:optimalmoments} indicates that the peaks of the PDFs change significantly with the addition of the fourth, fifth and sixth moments into the entropy calculation. Thus it appears that for the Schl\"{o}gl system we need at least six moments to accurately approximate the CDF of species $A$. The diagram on the right of Figure \ref{figure:optimalmoments} compares the empirical PDF and the approximate PDF using seven moments. Qualitatively, the approximate PDF captures the behaviour of the empirical PDF very well. However, in order to have a quantitative measure of the accuracy of approximate PDFs generated with different numbers of moments, we use the Kolmogorov-Smirnov distance \cite{sheskin-stats}.

Given two cumulative distribution functions (CDFs), $F(x)$ and $G(x)$, defined on support $\Omega$, the Kolmogorov-Smirnov distance is given by
\begin{equation}
\mathbb{D}_{F,G} \coloneqq \sup_{x \in \Omega} |F(x) - G(x)|.
\end{equation}
The Kolmogorov-Smirnov distance can be interpreted as the largest vertical distance between the two CDFs. The Kolmogorov-Smirnov distance between the approximate distribution calculated using the first seven moments and the empirical distribution is $0.0037$. As a comparison we define $c_i = \mathbb{E}\left[\mathbf{1}\{A(20) \leq i\}\right] = \mathbb{P}\left( A(20) \leq i \right)$ for $i \in \Omega$, and calculate the $95\%$ confidence intervals of each estimate $c_i$, the largest of which was $0.0001$. This means that the source of the error is in the maximum entropy method and is not a consequence of noise in the empirical CDF.

The multi-level method is used to generate the moments required by the maximum entropy method. In practice we do not want to rely on results from an eSSA to calculate the Kolmogorov-Smirnov distance so instead we assume that the PDFs on the left of Figure \ref{figure:optimalmoments} converge to the empirical distribution. We can therefore choose the number of moments, $i$, to be the minimal $i$ such that $ \mathbb{D}_{P_i,P_{i-1}} < \delta $, where $\delta$ is chosen to provide the desired level of accuracy and $P_i$ represents the CDF calculated using the first $i$ moments in the method of maximum entropy.

The question we now need to answer is which moment should have its estimator variance controlled when choosing the number of paths on each level in the multi-level method. Recall in Section \ref{section:modelling} we controlled the estimator variance of the first moment as this was the only system statistic of interest. However, now we have multiple system statistics. We choose to control the accuracy of every moment, $\mathbb{E}(X^k)$, by controlling its variance, $\mathrm{Var}\left(X^k \right)$, to within $\epsilon \cdot \mathbb{E} \left(X^k \right)$ for some $\epsilon > 0$. We now present an adaptive algorithm to generate an approximate CDF using the multi-level method together with the method of maximum entropy:

\begin{enumerate}
\item Set the number of levels $L$ and the number of initial moments $m \geq 2$. Set the variance control parameter, $\epsilon$, and the tolerance parameter for the Kolmogorov-Smirnov distance test, $\delta$.
\item Generate $1000$ initial sample paths on each level using the multi-level algorithm.
\item Calculate the number of additional paths, $n_{\ell}^j$, necessary on each level, $\ell$, to ensure $\mathrm{Var}\left(X^{j}\right) < \epsilon \cdot \mathbb{E}\left(X^{j}\right)$ using equation (\ref{equation:paths}) for each $j \in \{1,\ldots,m\}$. Find $n_{\ell}^{k_{\ell}} = \text{max}_{i \in \{1,\ldots,m\}} \left\lbrace n_{\ell}^i \right\rbrace$
\item Generate $n_{\ell}^{k_{\ell}}$ sample paths on each level, $\ell$, using the multi-level algorithm.
\item Let $P_1$ be the CDF generated using the maximum entropy method using $m-1$ moments, and $P_2$ the CDF using $m$ moments. If $\mathbb{D}_{P_1,P_2} < \delta$ terminate the algorithm, otherwise let $m=m+1$ and return to step $3$.
\end{enumerate}

This adaptive  algorithm systematically increases the number of moments used in the entropy calculation until the addition of a new moment has negligible effect on the resulting CDF, which is determined by the tolerance parameter, $\delta$. The algorithm relies on a convergence assumption, i.e. increasing the number of moments indefinitely will yield CDFs that converge to the true CDF. We note that, however, in practice, the algorithm is limited by the robustness of the numerical scheme chosen to calculate the CDFs. 

\begin{figure}[H]
\centering
$\begin{array}{cc}
\includegraphics[scale=1.1]{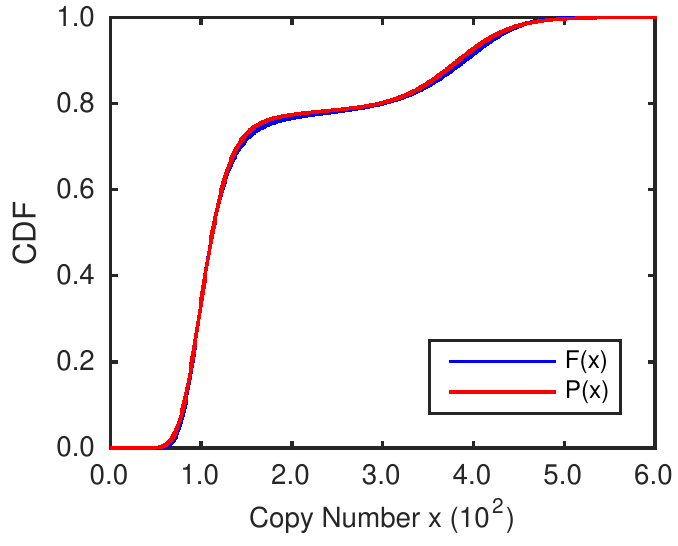} & \includegraphics[scale=1.1]{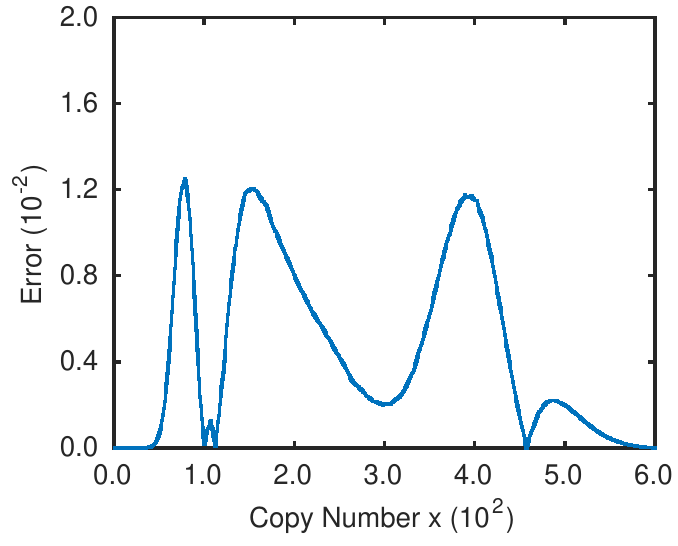}
\end{array}$
\caption{Left: The blue curve is the CDF $F(x)$ generated using Gillespie's DM with $10^6$ sample paths. The red curve is the CDF $P(x)$ generated using the adaptive multi-level algorithm and the method of maximum entropy, with control parameters $\epsilon=0.03$ and $\delta=0.02$. The truncated state space is $\Omega_t = \{0,1,2,\ldots,600\}$. The Kolmogorov-Smirnov distance is $0.0125$. Right: A plot of the error: $|F(x)-P(x)|$.}
\label{figure:schlogl:adaptive}
\end{figure}

In Figure \ref{figure:schlogl:adaptive} we present for the Schl\"{o}gl model: the CDF, $P(x)$, generated using the adaptive algorithm with $\epsilon=0.03$ and $\delta=0.02$; and the empirical CDF, $F(x)$, generated using Gillespie's DM which took $4568$ seconds and $38853$ seconds of CPU time to compute, respectively. The adaptive algorithm selected six as the optimal number of moments.

\subsubsection{A dimerisation model}

As our next example we consider gene expression for a single gene \cite{Anderson_Higham_12} $(G)$. It transcribes mRNA, known as messengers $(M)$, which can, in turn, translate proteins $(P)$. Pairs of these proteins may combine to form a dimer $(D)$. The mRNA and proteins may also degrade. This biochemical network is described by the following set of reactions \cite{Anderson_Higham_12}:
\begin{gather}
\begin{split}
R_1 : G \xrightarrow{k_1} G + M;\hspace{10 mm} R_2 : M \xrightarrow{k_2} M + P;\hspace{10 mm} R_3 : P + P \xrightarrow{k_3} D; \\
R_4 : M \xrightarrow{k_4} \emptyset; \hspace{10 mm} R_5 : P \xrightarrow{k_5} \emptyset. \hspace{30 mm}
\end{split} 
\label{equation:reactions:dimerisation}
\end{gather} 
We fix the rates to be $[k_1,k_2,k_3,k_4,k_5] = [25,1000,0.001,0.1,1]$ for the remainder of this paper. We wish to approximate the distribution of the dimer population at a fixed terminal time $T=1$ and we take the initial conditions to be $[G(0),M(0),P(0),D(0)]^T=[1,0,0,0]^T$.

\begin{figure}[H]
\centering
$\begin{array}{cc}
\includegraphics[scale=1.1]{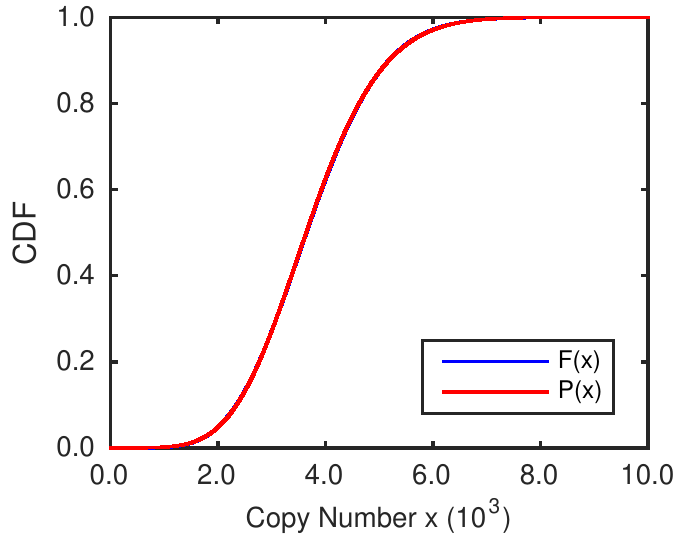} & \includegraphics[scale=1.1]{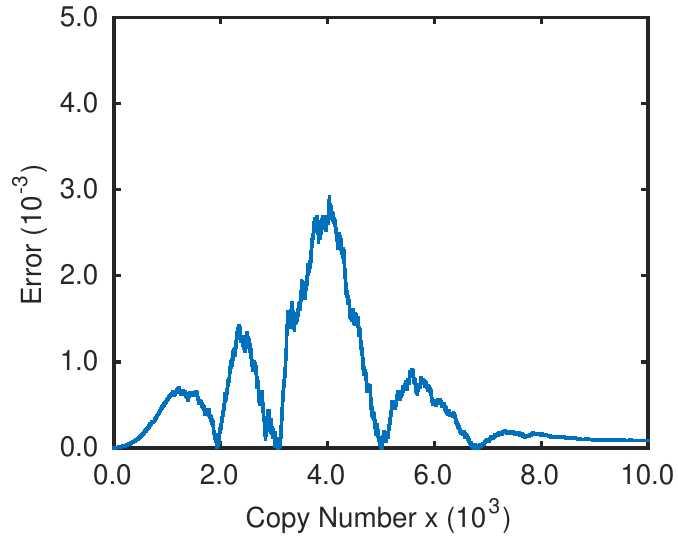}
\end{array}$
\caption{Left: The blue curve is the CDF $F(x)$ generated using Gillespie's DM with $10^6$ sample paths. The red curve is the CDF $P(x)$ generated using the adaptive multi-level algorithm and the method of maximum entropy, with control parameters $\epsilon = 0.0025$ and $\delta = 0.01$. The truncated state space is $\Omega_t = \{0, 1, 2, \ldots, 10000\}$. The Kolmogorov-Smirnov distance is $0.0029$. Right: A plot of the error: $|F(x)-P(x)|$.}
\label{figure:adaptive:ml:dim}
\end{figure}

Figure \ref{figure:adaptive:ml:dim} shows the CDF, $P(x)$, produced using the adaptive multi-level algorithm, it took $321$ seconds of CPU time to compute. The figure also contains the empirical CDF, $F(x)$, generated using Gillespie's DM, which took $2598$ seconds of CPU time to compute. The maximum $95\%$ confidence interval for the estimates $c_i$, where $c_i = \mathbb{E}[\mathbf{1}\{D(1) \leq i\}]$ for $i \in \Omega$, calculated using $10^6$ realisations of Gillepsie's DM was $0.0001$. This confirms that the error lies in the maximum entropy method and is not a consequence of noise in the empirical distribution. The adaptive algorithm selected four as the optimal number of moments.

\begin{figure}[H]
\centering
$\begin{array}{cc}
\includegraphics[scale=1.1]{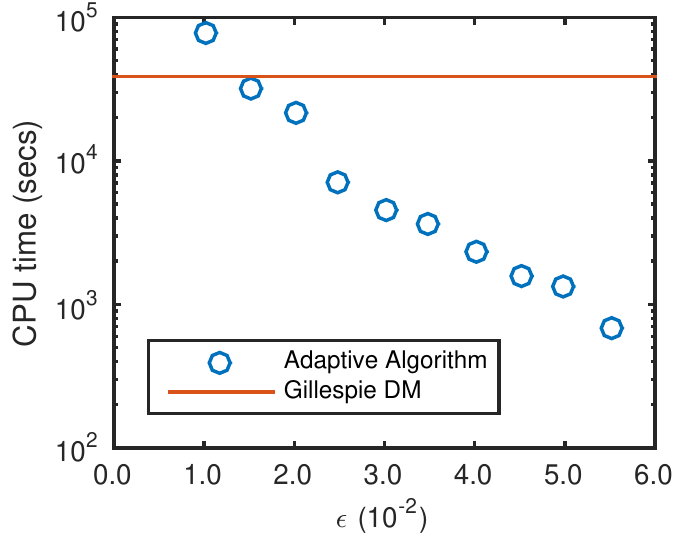} & \includegraphics[scale=1.1]{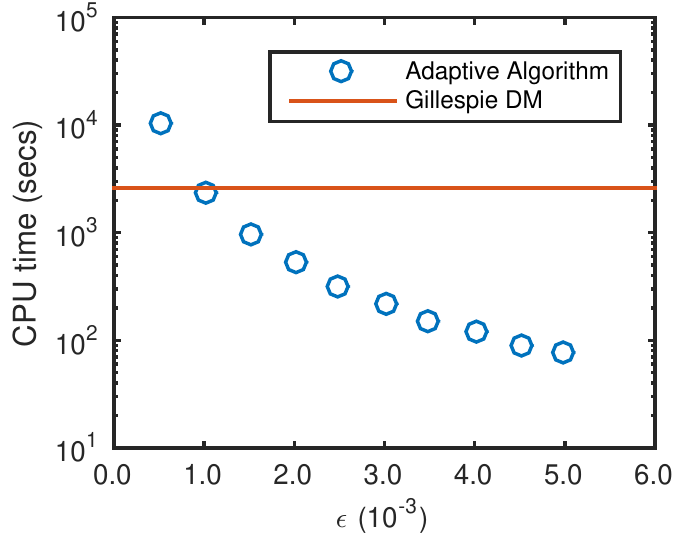}
\end{array}$
\caption{Left: CPU times for the adaptive algorithm using the method of maximum entropy for the Schl\"{o}gl model where $\delta = 0.02$. Right: CPU times for the adaptive algorithm using the method of maximum entropy for the dimerisation model where $\delta = 0.01$. In both plots the red line shows the CPU for $10^6$ sample paths from Gillespie's DM.}
\label{figure:maxent:CPU}
\end{figure}

Figure \ref{figure:maxent:CPU} shows how the CPU times increase for both the Schl\"{o}gl model and the dimerisation model as the variance control parameter, $\epsilon$, is decreased. The CPU times for the adaptive algorithm using the method of maximum entropy for the Schl\"{o}gl model are seen to jump at around $\epsilon=0.02$. This is because as $\epsilon$ is decreased the number of moments necessary to pass the Kolmogorov-Smirnov distance test increases, which results in a significant increase in the CPU time.

\subsection{Using an indicator function approach}

We now consider another method to approximate CDFs. Let $X$ be a distribution on a discrete state space $\Omega$. We can approximate the CDF of $X$ directly by estimating the following statistics,
\begin{equation}
c_i = \mathbb{P}\left( X \leq i \right) = \mathbb{E}\left[ \mathbf{1}\left\lbrace X \leq i \right\rbrace \right],
\end{equation}
for $i \in \Omega$. Although we refer to the collection of estimates $c_i$, for $i \in \Omega$, as an approximate CDF it is important to note that this method does not always generate a true CDF because estimates are not necessarily monotonic increasing i.e. $c_i \leq c_{i+1}$ for $\{i,i+1\} \in \Omega$. For two solutions to this problem see Section \ref{section:discussion}. 

These statistics, $c_i$, can be estimated using the multi-level method for which we control the variance of each of the $c_i$, requiring, as before, $\mathrm{Var}(\mathbf{1}\{ X \leq i \}) < \epsilon$ for some $\epsilon > 0$. One option is then to generate $N_{\ell} = \text{max}_{i \in \Omega} \{n_{\ell}^i \}$ paths on each level, where $n_{\ell}^i$ is the number of paths required on the $\ell^{th}$ level to control the variance of $c_i$. However, this requires large CPU times. Instead we introduce $I$ where $I \subset \Omega$. We generate $\hat{N_{\ell}} = \text{max}_{i \in I} \left\lbrace n_{\ell}^i \right\rbrace$ on each level and then calculate the estimates, $c_i$, for $i \in \Omega$. Then we iteratively increase the size of the subset $I$ until further increases have only negligible effects on the approximate CDFs produced. We update $I$ by finding the maximal set $J \subset I$ where each element $j \in J$ corresponds to the estimate $c_j$ that required the largest number of paths on some level, $\ell$, of the multi-level method. Then, for each $j \in J$, we add new entries to $I$ as the midpoints of $j$ and the nearest neighbours of $j$ that are already in $I$. We now present an algorithm for the adaptive method and the indicator function approach:

\begin{enumerate}
\item Set the number of levels L and the variance tolerance parameter, $\epsilon$, for all the estimates $c_i$. Set the tolerance parameter, $\delta$, for the Kolmogorov-Smirnov distance test. Set the initial subset $I \subset \Omega$ from which the number of paths to be generated will be decided.

\item Generate $1000$ initial sample paths on each level using the multi-level method.

\item Calculate the number of additional paths, $n_{\ell}^i$, necessary on each level, $\ell$, to ensure $\mathrm{Var}\left( \mathbf{1}\left\lbrace X \leq i \right\rbrace \right) < \epsilon$ for each $i \in I$ using equation (\ref{equation:paths}).

\item Generate $\hat{N_{\ell}}$ paths on each level, $\ell$, where $\hat{N_{\ell}}$ is given by,
\begin{equation}
\hat{N_{\ell}} = \max_{i \in I} \left\lbrace n_{\ell}^i \right\rbrace,
\end{equation}
using the multi-level method.

\item Let $P_1$ be the approximate CDF generated by estimating $c_i$ for each $i \in \Omega$.

\item Construct the maximal set $J$ such that for every $j \in J$, $n_{\ell}^j = \hat{N_{\ell}}$ for some level $\ell$. Then for every $j \in J$ add additional entries to $I$ as the midpoints between $j$ and its direct neighbours. If $I$ cannot be updated further, i.e. there is no midpoint not already in $I$, then terminate the algorithm.

\item Calculate the number of additional paths, $n_{\ell}^i$, necessary on each level $\ell$ to ensure $\mathrm{Var}\left( \mathbf{1}\left\lbrace X \leq i \right\rbrace \right) < \epsilon$, for each $i \in I$ using equation (\ref{equation:paths}).

\item Generate $\hat{N_{\ell}}$ paths on each level, $\ell$, where $\hat{N_{\ell}}$ is given by,
\begin{equation}
\hat{N_{\ell}} = \max_{i \in I} \left\lbrace n_{\ell}^i \right\rbrace,
\end{equation}
using the multi-level method.

\item Let $P_2$ be the approximate CDF created by calculating $c_i$ for each $i \in \Omega$.

\item If $\mathbb{D}_{P_1,P_2} < \delta$, terminate the algorithm, otherwise let $P_1 = P_2$ and return to step $6$.
\end{enumerate}

We note that $c_i \in [0,1]$ for every $i \in \Omega$ and so we do not scale the variance tolerance of each estimate by $c_i$, instead we select a single variance tolerance, $\epsilon$, for all estimates. We now use this adaptive algorithm on the two example systems we have seen previously.

\subsubsection{The Schl\"{o}gl model}

\begin{figure}
\centering
$\begin{array}{cc}
\includegraphics[scale=1.1]{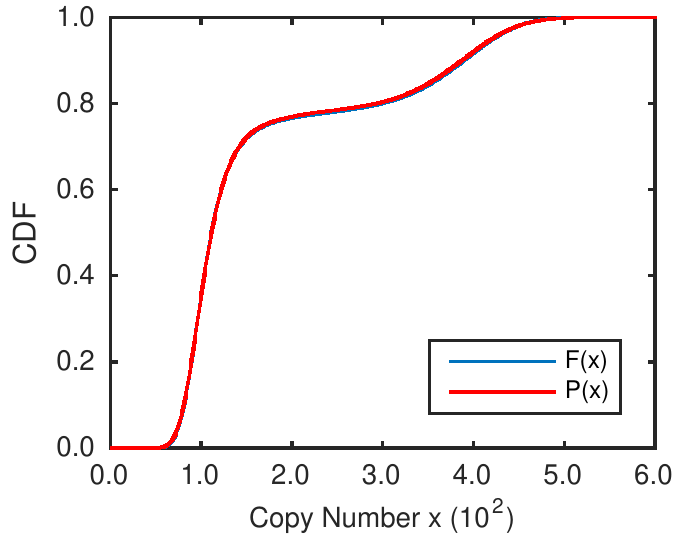} & \includegraphics[scale=1.1]{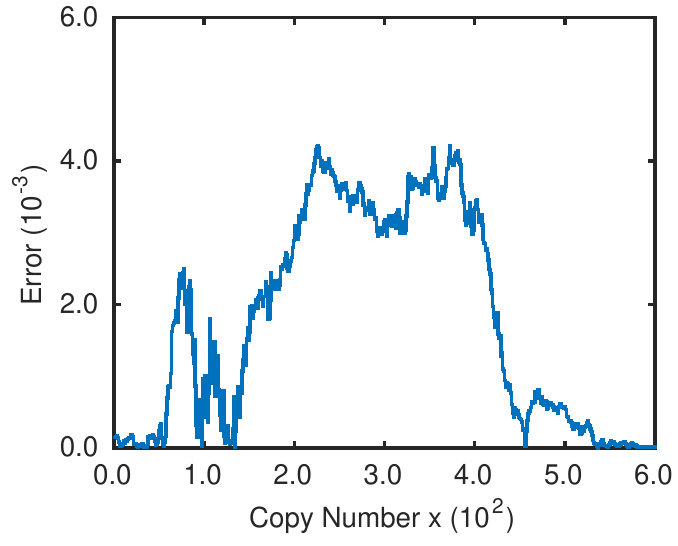}
\end{array}$
\caption{Left: The blue curve is the CDF $F(x)$ generated using Gillespie's DM with $10^6$ sample paths. The red curve is the approximate CDF $P(x)$ generated using the adaptive multi-level algorithm for the indicator function approach, with control parameters $\epsilon = 0.0025$ and $\delta = 0.001$. The truncated state space is $\Omega_t = \{0,1,2,\ldots,600\}$. The Kolmogorov-Smirnov distance is $0.0042$. Right: A plot of the error: $|F(x)-P(x)|$.}
\label{figure:adapt:int:schlogl}
\end{figure}

We consider the molecular species $A$ and the reactions (\ref{equation:schlogl:reactions}). We truncate the state space to $\Omega_t = \{0,\ldots,600\}$ and consider the initial condition $A(0)=0$ and terminal time $T=20$. The initial subspace used is $I = \{100, 200, 300, 400, 500\}$. In Figure \ref{figure:adapt:int:schlogl} we present the approximate CDF $P(x)$ generated using the adaptive multi-level algorithm with $\epsilon = 0.0025$ and $\delta = 0.001$ and the CDF $F(x)$ generated using Gillespie's DM which took $6762$ seconds and $38853$ seconds of CPU time, respectively. The CDFs are very similar and the exact error is given on the right of Figure \ref{figure:adapt:int:schlogl}.

\subsubsection{A dimerisation model}

\begin{figure}
\centering
$\begin{array}{cc}
\includegraphics[scale=1.1]{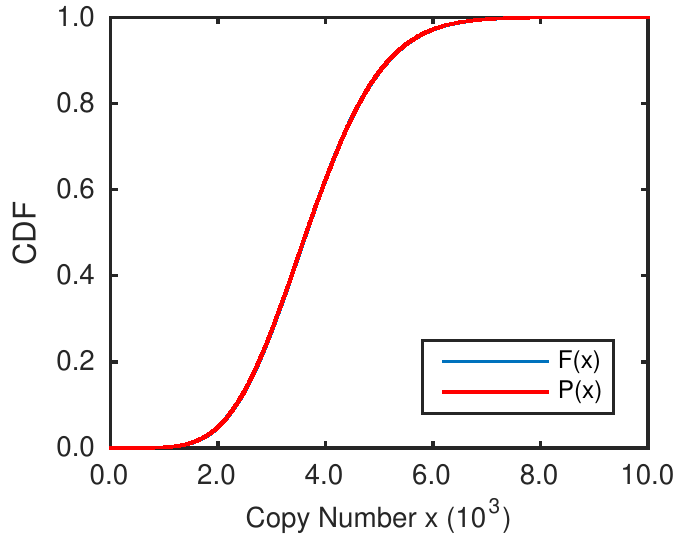} & \includegraphics[scale=1.1]{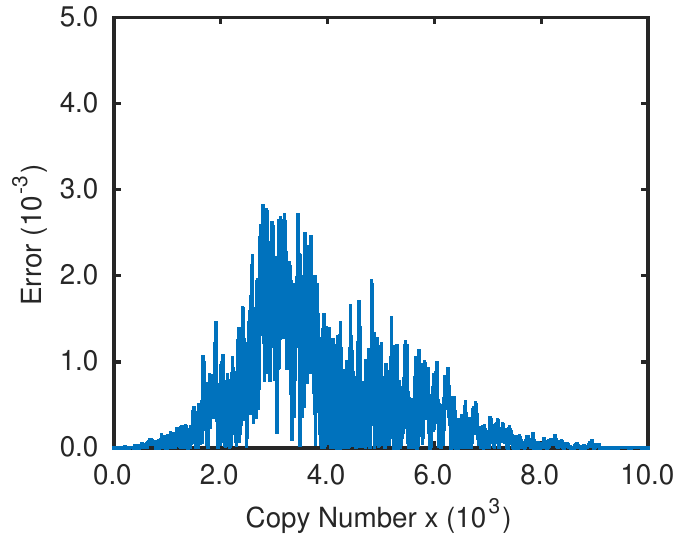}
\end{array}$
\caption{Left: The blue curve is the CDF $F(x)$ generated using Gillespie's DM with $10^6$ sample paths. The red curve is the approximate CDF $P(x)$ generated using the adaptive multi-level algorithm for the indicator function approach, with control parameters $\epsilon = 0.002$ and $\delta = 0.001$. The truncated state space is $\Omega_t = \{0,1,2,\ldots,10000\}$. The Kolmogorov-Smirnov distance is $0.0028$. Right: A plot of the error: $|F(x)-P(x)|$.}
\label{figure:adapt:int:dim}
\end{figure}

Consider the four species $G$, $M$, $P$ and $D$ and the reactions (\ref{equation:reactions:dimerisation}). We wish to approximate the distribution of the dimer population at a fixed time time $T=1$ where we take the initial conditions to be $[G(0),M(0),P(0),D(0)]^T=[1,0,0,0]^T$. We truncate the state space to $\Omega_t = \{0,\ldots,10000\}$ and the initial subspace is $I = \{1000,2000,\ldots,8000,9000\}$. In Figure \ref{figure:adapt:int:dim} we present the approximate CDF $P(x)$ generated using the adaptive multi-level algorithm with $\epsilon = 0.002$ and $\delta = 0.001$ and the CDF $F(x)$ generated using Gillespie's DM which took $249$ seconds and $2598$ seconds of CPU time, respectively. Again the CDFs are very similar and the exact error is given on the right of Figure \ref{figure:adapt:int:dim}.

\begin{figure}
\centering
$\begin{array}{cc}
\includegraphics[scale=1.1]{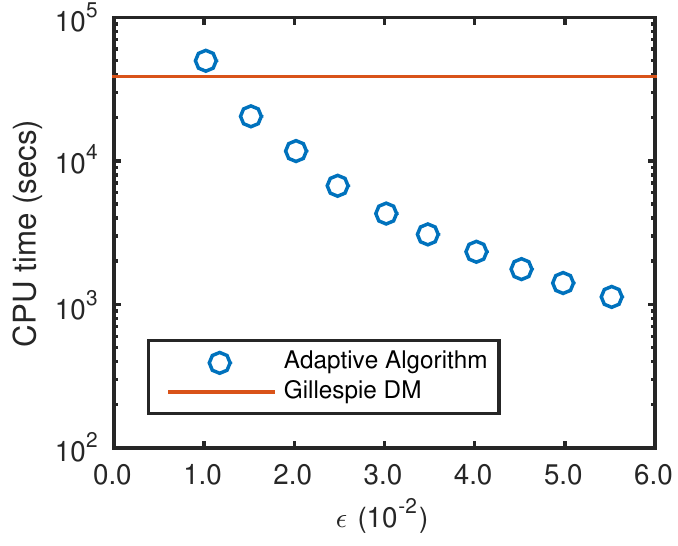} & \includegraphics[scale=1.1]{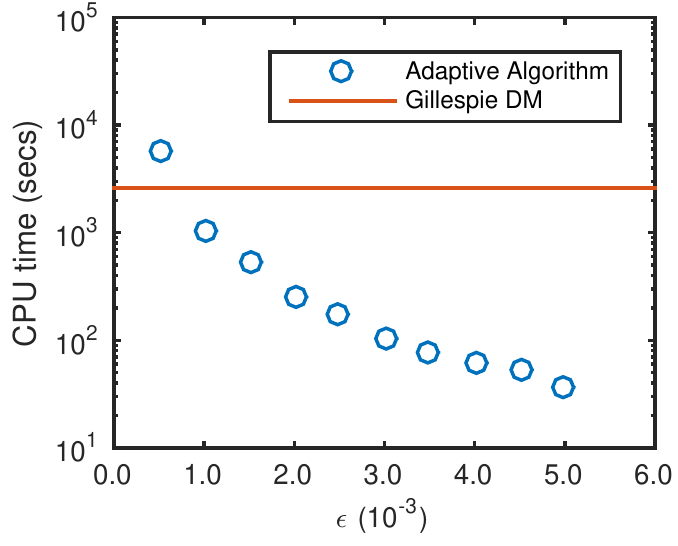}
\end{array}$
\caption{Left: CPU times for the adaptive algorithm using the indicator function approach for the Schl\"{o}gl model where $\delta = 0.001$. Right: CPU times for the adaptive algorithm using the indicator function approach for the dimerisation model where $\delta = 0.001$. In both plots the red line shows the CPU for $10^6$ sample paths from Gillespie's DM.}
\label{figure:CPU:direct}
\end{figure}

Figure \ref{figure:CPU:direct} shows how the CPU times increase for both the Schl\"{o}gl model and the dimerisation model as the variance control parameter, $\epsilon$, is decreased. As $\epsilon$ is decreased we see that the CPU time of the adaptive method and the indicator function approach becomes larger than the CPU time for Gillespie's DM.

\section{Discussion} \label{section:discussion}

We have presented two novel algorithms designed to employ the multi-level Monte Carlo method to generate approximate CDFs for molecular populations at a terminal time $T$. The first was the method of maximum entropy which used the multi-level method to generate a  finite set of moments and from these moments calculated an approximate distribution. We note that this method is favourable if an analytical approximation of the PDF is required. The second method was an indicator function approach. This adaptive method focussed on generating more paths to control the estimates of $c_i$ in regions of $\Omega$ where the random variables $\mathbf{1}\{X\leq i\}$ have a higher variance. These regions are often found to be near the peaks of the corresponding PDFs which is where the method of maximum entropy returns the largest errors. Thus we see that the indicator function approach achieves smaller Kolmogorov-Smirnov distances than the method of maximum entropy when using the Schl\"{o}gl model. 

However, we note that the indicator function approach does not necessarily provide a CDF as the estimates $c_i$ are not always monotonic increasing in $i$. Although the estimates can still be used to give an accurate and meaningful description of the distribution, if a physical CDF is required there are at least two solutions. Firstly one can vastly decrease the value of the variance control parameter $\epsilon$ to reduce the probability of non-montonicity, this however will significantly increase the CPU time of the method such that it is comparable with the tau-leap method and Gillespie's DM. Secondly one can take estimates for $c_i$ from a coarse subset $S \subset \Omega$, and generate the CDF using polynomial interpolation whilst enforcing a positive gradient over the range of $\Omega$. This second approach uses a coarser subset $S$ to reduce the probability of non-monotonicity without the further increase in the CPU time.  

Future work will involve investigating the potential for improvements on the efficiency of the maximum entropy method, as well as improvements in the efficiency of using multi-level methods to produce approximate CDFs in general.

\section*{Acknowledgements}

The authors would like to thank Chris Lester for many insightful discussions.

\addcontentsline{toc}{section}{References}
\bibliographystyle{abbrv}
\interlinepenalty=10000
\bibliography{finalpaperbib}

\begin{thebibliography}{10}

\bibitem{Anderson_Higham_12}
D.~F. Anderson and D.~J. Higham.
\newblock Multi-level {M}onte {C}arlo for continuous time {M}arkov chains with
  applications in biochemical kinetics.
\newblock {\em SIAM Multiscale Modeling and Simulation}, 10(1):146--179, 2012.

\bibitem{fluorescence}
M.~B. Elowitz, A.~J. Levine, E.~D. Siggia, and P.~S. Swain.
\newblock Stochastic gene expression in a single cell.
\newblock {\em Science}, 297(5584):1183--1186, 2002.

\bibitem{Engblom}
S.~Engblom.
\newblock Spectral approximation of solutions to the chemical master equation.
\newblock {\em Journal of Computational and Applied Mathematics},
  229(1):208:221, 2009.

\bibitem{SNIPER}
R.~Erban, J.~S. Chapman, I.~G. Kevrekidis, and T.~Vejchodsk{\'y}.
\newblock Analysis of a stochastic chemical system close to a sniper
  bifurcation of its mean-field model.
\newblock {\em SIAM Journal of Applied Mathematics}, 70(3):984:1016, 2009.

\bibitem{Radek-guide}
R.~Erban, J.~S. Chapman, and P.~K. Maini.
\newblock A practical guide to stochastic simulations of reaction-diffusion
  processes.
\newblock {\em arXiv}, 0704(1908), 2007.

\bibitem{Giles}
M.~B. Giles.
\newblock Multilevel {M}onte {C}arlo path simulation.
\newblock {\em Operations Research}, 56(3):607--617, 2008.

\bibitem{Gillespie}
D.~T. Gillespie.
\newblock Exact stochastic simulation of coupled chemical reactions.
\newblock {\em The Journal of Physical Chemistry}, 81(25):2340--2361, 1977.

\bibitem{tau-leap}
D.~T. Gillespie.
\newblock Approximate accelerated stochastic simulation of chemically reacting
  systems.
\newblock {\em Journal of Chemical Physics}, 115(4):1716--1733, 2001.

\bibitem{Jahnke_2}
T.~Jahnke.
\newblock On reduced models for the chemical master equation.
\newblock {\em Multiscale Modelling and Simulation}, 9(4):1646--1676, 2011.

\bibitem{Jahnke}
T.~Jahnke and W.~Huisinga.
\newblock Solving the chemical master equation for monomolecular reaction
  systems analytically.
\newblock {\em Journal of Mathematical Biology}, 54(1):1--26, 2007.

\bibitem{jaynes}
E.~T. Jaynes.
\newblock {\em Probability Theory: The Logic of Science}.
\newblock Cambridge University Press, 2003.

\bibitem{Kurtz-2011}
T.~G. Kurtz and D.~F. Anderson.
\newblock Continuous time {M}arkov chain models for chemical reaction networks.
\newblock In {\em Design and Analysis of Biomolecular Circuits}, pages 3--42.
  Springer, 2011.

\bibitem{userguide}
C.~Lester, R.~E. Baker, M.~B. Giles, and C.~A. Yates.
\newblock A guide to efficient discrete-state multi-level simulation of
  stochastic biological systems.
\newblock {\em arXiv}, 1412(4069), 2014.

\bibitem{adaptive-lester}
C.~Lester, C.~A. Yates, M.~B. Giles, and R.~E. Baker.
\newblock An adaptive multi-level simulation algorithm for stochastic
  biological systems.
\newblock {\em The Journal of Chemical Physics}, 142(2):024113, 2015.

\bibitem{matlab-entropy}
A.~Mohammad-Djafari.
\newblock A {M}atlab program to calculate maximum entropy distributions.
\newblock {\em Fundamental Theories of Physics}, 50:221--233, 1992.

\bibitem{schlogl_origin}
F.~Schl{\"o}gl.
\newblock Chemical reaction models for non-equilibrium phase transitions.
\newblock {\em Zeitschrift f{\"u}r Physik}, 253(2):147:161, 1972.

\bibitem{sheskin-stats}
D.~J. Sheskin.
\newblock {\em Handbook of {P}arametric and {N}on-parametric {S}tatistical
  {P}rocedures}.
\newblock Chapman and Hall, third edition, 2003.

\bibitem{Calculus}
J.~Stewart.
\newblock {\em Calculus}.
\newblock Brooks/Cole, sixth edition, 2009.

\bibitem{suli}
E.~S{\"u}li and D.~F. Mayers.
\newblock {\em An Introduction to Numerical Analysis}.
\newblock Cambridge University Press, 2003.

\bibitem{Szekely_2012}
T.~Sz{\'e}kely, K.~Burrage, and R.~Erban.
\newblock A higher order numerical framework for stochastic simulation of
  chemical reaction systems.
\newblock {\em BMC Systems Biology}, 6(1):85, 2012.

\bibitem{schlogl}
T.~Wilhelm.
\newblock The smallest chemical reaction system with bistability.
\newblock {\em BMC Systems Biology}, 3(1):90, 2009.

\end{thebibliography}
\end{document}